\documentclass[pra,twocolumn,showpacs,preprintnumbers,amsmath,amssymb]{revtex4}
\makeatletter
\usepackage[dvips]{graphicx}	

\usepackage{amsmath}
\usepackage{amssymb}
\usepackage{amsbsy}
\usepackage{color}
\setcitestyle{square}
\usepackage{epstopdf}
\DeclareGraphicsExtensions{.pdf,.eps,.png,.jpg,.mps}

\begin{document}

\title{
Gapless states and current control in strongly distorted gated trilayer graphene }
\author{ W. Jask\'olski} \email{wj@fizyka.umk.pl}
\affiliation{Institute of Physics, Faculty of Physics, Astronomy and Informatics, Nicolaus Copernicus University in Torun, Grudziadzka 5, 87-100 Toru\'n, Poland}

\begin{abstract}
We investigate gated trilayer graphene partially devoid of outer layers and forming a system of two trilayers connected by a single layer of graphene. A difference in the stacking order of trilayers leads to the appearance of gapless states, one of which is mainly localized in the single graphene layer. We demonstrate that by changing the value of the gate voltage applied to the outer layers one can change the slope of $E(k)$ of this state. As a consequence the direction of current flowing in the single layer graphene can also be changed, the effect that could be useful in practical applications.
\end{abstract}

\date{\today}

\maketitle

\section{Introduction}

There are several reasons for the high interest in multilayer graphene. One of them is the possibility of opening a tunable energy gap in gated multilayers with the Bernal-stacked arrangement of layers, which is important for electronic applications of graphene \cite{Ohta_2006, Castro_2007,Oostinga_2008,Zhang_Nature_2009,Szafranek_2010,Padilha_2011,Schwierz_2010,Lin_2008,Choi_2010,Santos_2012,Zhang_transistor_2018}. Multilayers focus also attention due to the emergence of flat bands and superconductive properties discovered for systems with twisted layers \cite{Jarillo2018,Chen_Nature_2019,Chittari_PRL_2019,Yin_PRB_2020, Liu_Nature_2020,Shen_Nat_Phys_2020}. Vacancies, adatoms and topological defects in multilayers are also investigated because they can induce magnetic moments important for spintronic applications \cite{An_SN_2018, Kishimoto_2016,Menezes_JPB_2015,Jaskolski_2016,Jaskolski_RSC_2019,Yazyev_RPP_2010,Nair_NC_2013}. Of particular interest are multilayers with domain walls that separate  regions of different stacking order, due to  the appearance of valley-protected gap states of topological character \cite{Wu_ACS_Nano_2015,Yin_NC_2016,
Kazemi_APL_2013,Anderson_PRB_2022,Vaezi_2013,Alden_2013,San_Jose_2014,Pelc_2015, Jaskolski_2020}. Two gap states appear in gated bilayer graphene (BLG) when the stacking order changes from AB to BA, and three such states occur in the energy gap of gated trilayer graphene (TLG) at stacking domain wall ABC/CBA. 

Recently it has been shown that strong deformation of gated bilayer and trilayer graphene with stacking domain walls, caused by partial removal of one of the layers, does not destroy their basic electronic properties, 
in particular topologically protected states in the energy gap \cite{Jaskolski_2019,Jaskolski_2021}. Here we investigate trilayer graphene partially peeled off the outer layers and thus forming a system of two trilayers connected by a single layer of graphene (SLG). Positive and negative $\pm V$ voltages are applied to outer layers. When the trilayers differ in the stacking arrangement (ABC and CBA), valley-protected gapless states are still present in the energy gap although their number depends on the gate voltage $V$. Some of these states localize almost exclusively in the region of single graphene layer  providing one-dimensional currents in this layer. Most importantly, by changing the value of $V$ the slope of $E(k$) of the gapless state central to the energy cone also changes. In practice, this means the change of the direction of current that flows in the SLG, the effect that could be exploited in nanoelectronic devices based on multilayer graphene.

\begin{figure}[ht]
\centering
\includegraphics[width=\columnwidth]{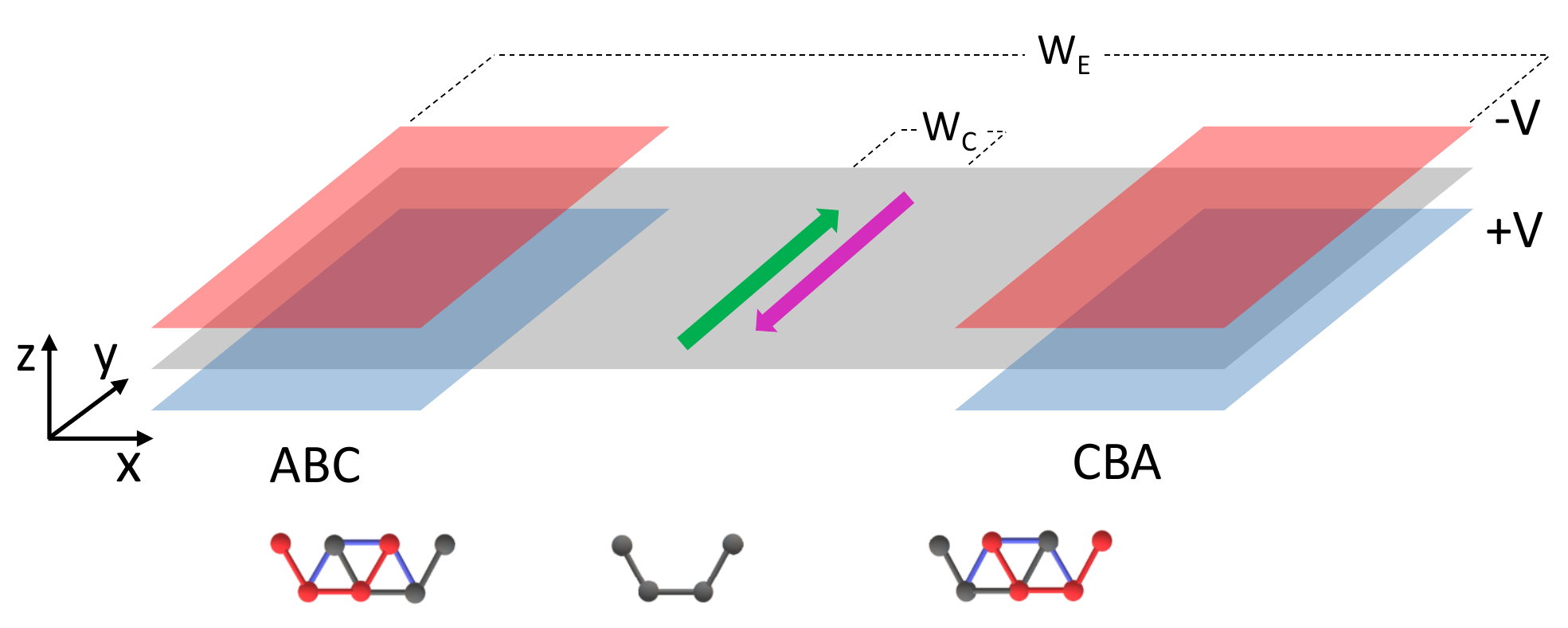}
\caption{\label{fig:first}
Schematic representation of the investigated TLG/SLG/TLG system.
The left and right TLGs have ABC and CBA arrangement of layers, respectively. The system extends to infinity in the $x$ (armchair) and $y$ (zigzag) directions, but is fully periodic only in the $y$ direction.
The top layer is marked in red, middle layer in gray and bottom layer in blue. $W_E$ and $W_C$ mark the widths of regions where LDOS is calculated.
At the bottom the top view of twelve-atom unit cells of each trilayer and four-atom unit cell of the SLG is depicted. 
Only atoms from adjacent layers, that overlap in the figure, are connected by 
$\gamma_1$, which is red to grey and blue to gray. 
The green and ping arrows indicate the change of the direction of current flowing in the central part of the SLG when the value of $V$ changes. 
}
 \end{figure}

\section{System investigated and method of calculation}

The investigated system is schematically shown in Fig. \ref{fig:first}.
It consists of two trilayers connected by a single graphene layer, i.e., TLG/SLG/TLG. 
The width of the central single layer is $W_{SLG}=7$, measured as the number of four-atom unit cells along the $x$ (armchair) direction. It means that the TLGs are separated by about 3 nm. The system is infinite in both, the armchair and the zigzag directions, but is periodic only in the zigzag direction. The stacking order of layers in the left and right TLGs is ABC and CBA, respectively. 

We work in the $\pi$-electron tight binding approximation. Intra-layer and inter-layer hopping parameters $\gamma_0=2.7$ eV and $\gamma_1 = 0.27$ eV are used, respectively \cite{Castro_2007,Ohta_2006}. Voltages $-V$ and $+V$ are applied to the top and bottom layers, respectively.  
This keeps the Fermi level at zero energy. We consider two values of $V$, 
0.1 eV and 0.5 eV, because the layer localization of gapless states in multilayers with stacking domain walls depends on the value of $V$ vs $\gamma_1$ \cite{Jaskolski_2018,Jaskolski_2020}. Surface Green function matching technique (SGFM) for three-terminal device (TLG/SLG/TLG) is used to calculate the local density of states (LDOS) \cite{Nardelli_1999}.  

Since the system is periodic in the zigzag direction the LDOS is $k$-dependent, where $k$ is the wave-vector corresponding to this periodicity. The LDOS is calculated in two regions: (i) in the entire structure shown in Fig. \ref{fig:first} of the width $W_E=19$  
and (ii) in the very central part of SLG of the width $W_C=3$. 
It has to be emphasized that although the LDOS is computed for finite segments of the widths $W_E$ and $W_C$, the calculation within the SGFM takes into account the structure that extends to infinity on the left and right sides.
To determine the localization of the gapless states we calculate also  layer-resolved density distribution for individual states at the Fermi level i.e., in the center of the energy gap. 

It is worth noting that because both outer layers are torn into two 
well-separated parts we cannot clearly indicate the position of the domain wall, as is the case when some layers are stretched, folded, delaminated or with a defect line \cite{Pelc_2015,Ju_Nature_2015,Lin_NL_2013,Peeters_PRB_2018,Jaskolski_2021}. Moreover, the zigzag edges of the half-planes of outer layers introduce zigzag-edge states that are expected to influence the energy spectrum of the investigated system.

\begin{figure}[ht]
\centering
\includegraphics[width=\columnwidth]{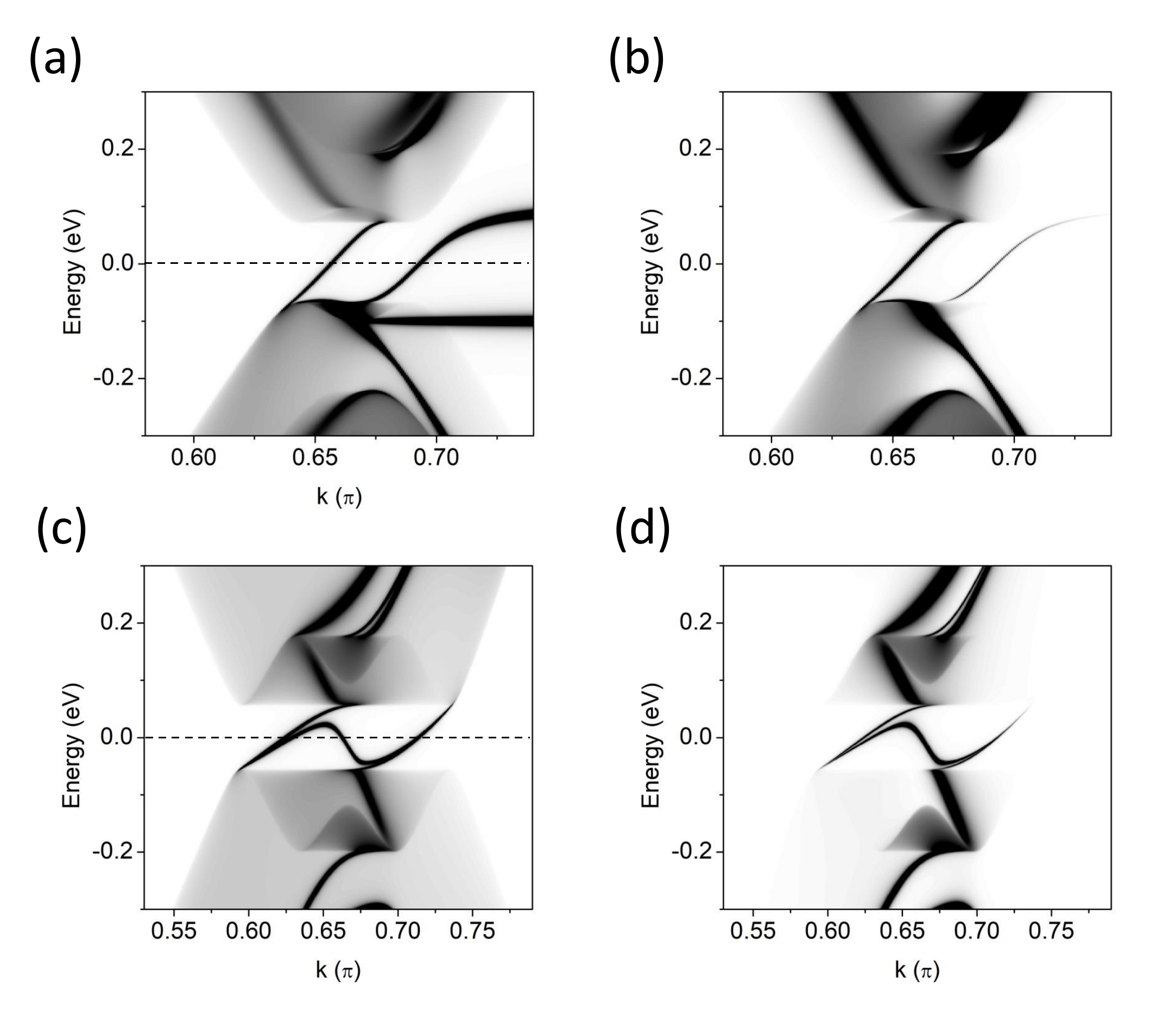}
\caption{\label{fig:second}
(a) and (c) LDOS calculated in the entire structure shown in Fig. \ref{fig:first}; (b) and (d) LDOS calculated in the central part of SLG; (a) and (b)  $V=0.1$ eV, (c) and (d)  $V=0.5$ eV. 
LDOS is visualized close to the energy cone, i.e., 
near the Fermi level 
and for $k$ around $\frac{2}{3} \pi $. 
In (c) and (d) the zigzag edge bands are at $E=\pm 0.5$ eV and are out of the figure range.
}
 \end{figure}

\section{Results and discussion}

First, let us recall that the energy spectrum of undisturbed, gated TLG with  ABC/CBA stacking domain wall presents an energy gap and a triplet of gap  states that connect the valence (VB) and conduction (CB) band continua \cite{Jaskolski_2020}. The slope of $E(k)$ of all gap states is the same and for a given valley and is uniquely determined by the gate polarization \cite{Jaskolski_2018}. As demonstrated in Ref. \cite{Jaskolski_2021} one state of this triplet localizes mainly in the lower layer, another one in the upper layer and the third one in both outer layers with an admixture in the middle layer.

\begin{figure}[ht]
\centering
\includegraphics[width=\columnwidth]{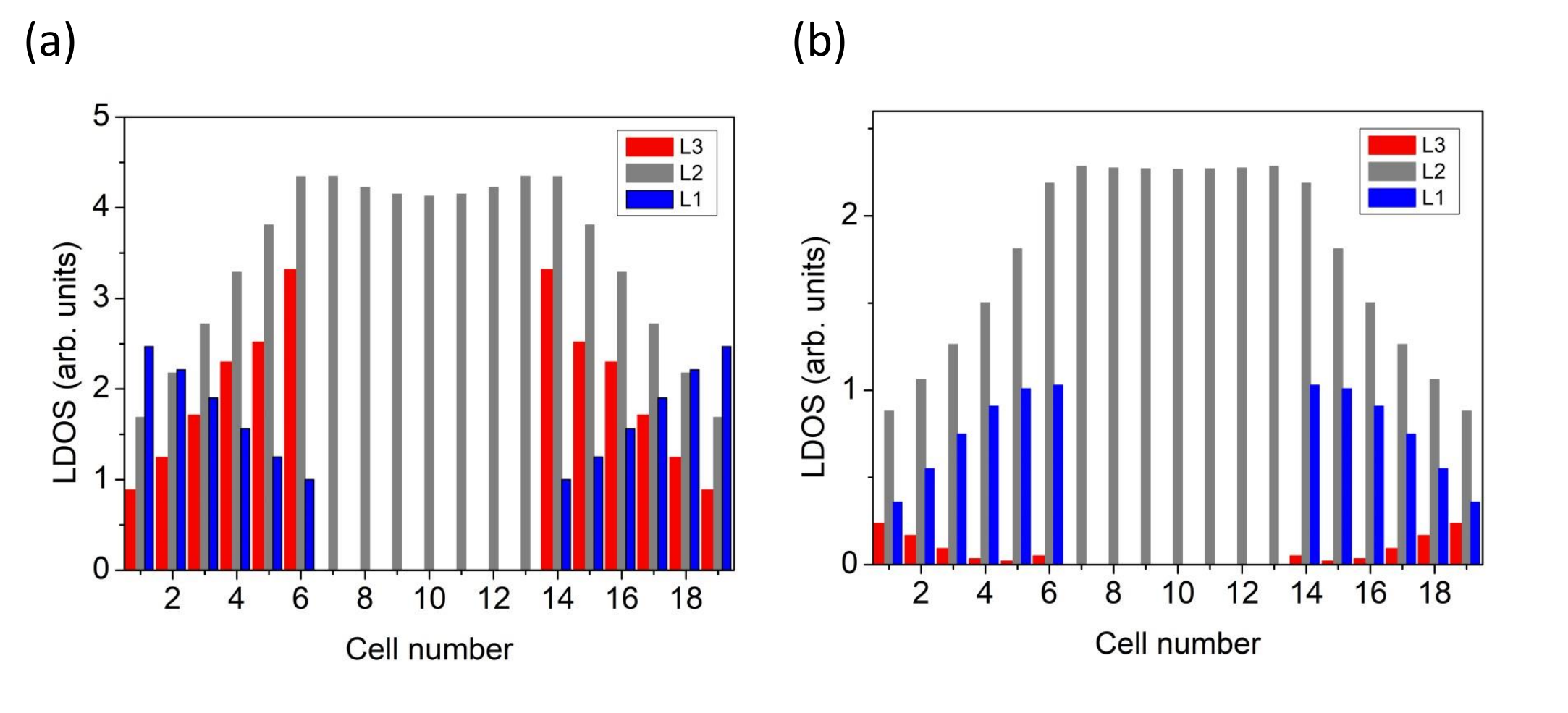}
\caption{\label{fig:third}
Layer-resolved density distribution of the gapless states 
shown in Fig. \ref{fig:third} (a) and (b) for $E=E_F$ and $k \approx \frac{2}{3} \pi $. 
(a)  the case of $V=0.1$ eV, (b) $V=0.5$ eV.  Vertical bars represent densities calculated at four-atom unit cells (as shown in Fig. \ref{fig:first}) in each layer. Lower layer (L1) - blue, middle layer (L2) - grey, upper layer (L3) - red. Six unit cells are taken into account in the left and right TLGs, so the SLG part 
extends from cell number 7 to cell number 13.
}
 \end{figure}

We start description of the results with the case $V=0.1$. In Fig. \ref{fig:second} (a) the LDOS calculated for the region of width $W_E$ is shown. Although the TLG is severely perturbed, i.e., strips of width $W_{SLG}$ are removed from outer layers and no stacking domain wall can be clearly identified, one state linking VB and CB still persists. For $E=E_F$ and $k \approx \frac{2}{3} \pi$ i.e., close to the cone center, this state is localized mainly in the SLG, what is confirmed in Fig \ref{fig:third} (a), where its layer-resolved density distribution is presented.
Two zigzag edge states are seen for $k\rightarrow \pi$ \cite{nota_zigzag_sublat}. The one at $E=-0.1$ eV is due to the zigzag edges of the upper half-planes of graphene \cite{nota_zigzag_degener}.
The zigzag-edge state at $E=+0.1$ eV, which is due to the zigzag edges of the lower half-planes, couples strongly to another gap state, which starts at the top of the valence band. 
It has been checked that for $E=E_F$ this state localizes mainly in the lower layer and its component in the middle layer is negligible
 
Fig. \ref{fig:second} (b) shows LDOS calculated in the central part of SLG of width $W_C$ (see Fig. \ref{fig:first}). The most striking effect is  strong resemblance of this energy spectrum to the LDOS of  the entire structure of width $W_E$, although the TLGs are distant from the $W_C$ region of SLG. It means that the energy spectrum of the TLGs is induced in the single layer of graphene. Consequently, it also means that the use of a single layer may be sufficient to take advantage of the properties of the entire system.

The zigzag-edge state at $E=-0.1$ eV is not seen because the sublattice of the upper layer at which this state localizes is decoupled from the middle layer. The zigzag-edge state at $E=0.1$ eV is only weakly represented, since its sublattice in the upper layer is weakly coupled (by $\gamma_1$) to the middle layer.

\begin{figure}[ht]
\centering
\includegraphics[width=\columnwidth]{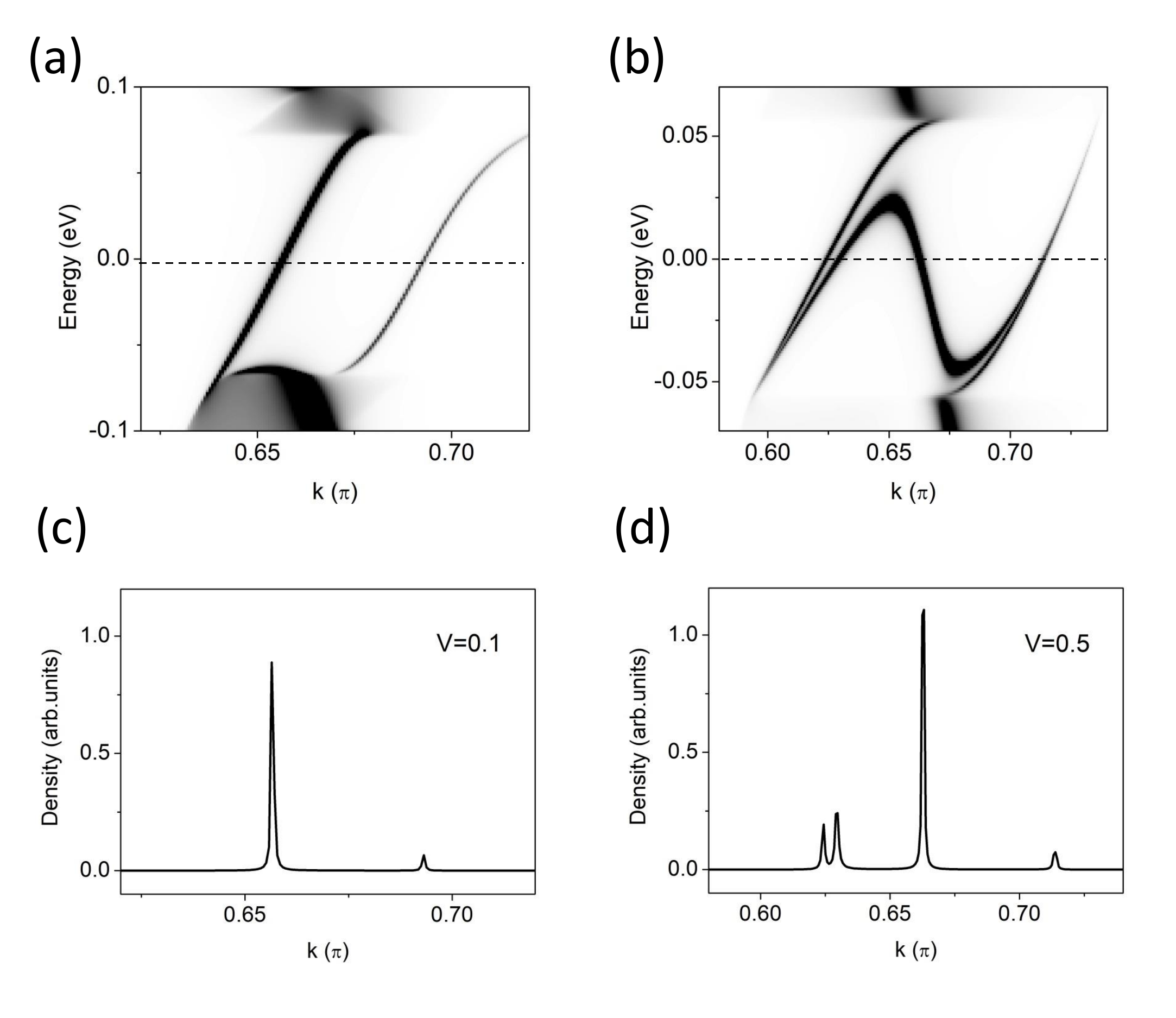}
\caption{\label{fig:fourth}
(a) and (b) magnification of the gapless states shown in Fig. \ref{fig:second} (b) and (d), i.e., for $V=0.1$ eV and $V=0.5$ eV, respectively. Horizontal broken line marks the Fermi level, for which the  cut of the LDOS is visualized in (c) and (d). 
}
 \end{figure}

The energy spectrum of the investigated system for $V=\pm 0.5$ eV is shown in Fig. \ref{fig:second} (c) \cite{nota_z_05}. For this gate voltage, three gapless states are maintained, although they differ significantly from the triplet of gap states that occur in the case of TLG with ABC/CBA stacking domain wall \cite{Jaskolski_2021}. 
Here, the middle gap state changes twice the slope of $E(k)$. To see better this effect the magnification of the energy spectrum close to $E_F$ is presented in Fig. \ref{fig:fourth} (b).
We have checked that the left and right gap states are localized almost exclusively in the outer layers. However, the middle gap state, similarly to the gap state of the $V=0.1$ case, localizes mainly in the middle layer, as shown in Fig. \ref{fig:third} (b).
There is a certain correspondence between these two states: both cross the Fermi level at $k \approx \frac{2}{3} \pi$ (i.e. at the cone center) and both localize in the middle layer.

The LDOS calculated in the central part of SLG is presented in Fig. \ref{fig:second} (d). Similar to $V=0.1$ eV, also this time the energy spectrum of the entire system is, to a large extent, induced in the single graphene layer. The topological gap states are well seen and the band continua are also partially preserved. 

Magnification of the LDOS close to the cone center, calculated in the central part of the SLG, for $V=0.1$ eV and $V=0.5$ eV are presented in Fig. \ref{fig:fourth} (a) and (b), respectively. The LDOS calculated at $E=E_F$ is presented in Figs. \ref{fig:fourth} (c) and (d). This figure   
shows that the main contribution to charge density at $E=E_F$ 
comes from the states that cross $E_F$ at $k \approx \frac{2}{3} \pi$.

These gap states can curry one dimensional currents along the zigzag direction, localized in the SLG. For $k$ in the adjacent valley, i.e., $k=-\frac{2}{3} \pi$, all the gap states have reversed slopes due to time-reversal symmetry, so the currents are valley-polarized. This property was proposed long ago to use as {\it valley filter} and {\it valley valve} \cite{Martin_PRL_2008} in valleytronics \cite{Rycerz_2007,Kundu_2016}. 
Polarization of valley currents can be changed by the reversal of the gate voltage \cite{Jaskolski_2016}. 
Here we demonstrate that for the system under study also the slope  of $E(k)$ of the gap state at $E_F$ and $k=\frac{2}{3} \pi$ changes with the change of magnitude of $V$. Therefore, we gain the possibility of changing the direction of valley current in this state by changing only the value of the gate voltage, what in Fig. \ref{fig:first} is illustrated by green and pink arrows.

\section{Conclusions}

We have studied gated trilayer graphene, strongly distorted by removal of wide strips of the outer layers. The resulting system consists of two trilayers of different stacking order, ABC and CBA, connected by single graphene layer. 
We have investigated the energy spectrum of the entire TLG/SLG/TLG and the local density of states in the center of  SLG. The calculations have been performed for two values, 0.1 eV and 0.5 eV, of the gate voltage applied to outer layers of TLGs.

The ABC/CBA stacking domain wall in gated trilayer graphene causes the emergence of three topologically protected states in the energy gap. Our calculations show that when such a trilayer is strongly deformed and  reduced to the TLG/SLG/TLG system, some gapless state still exist, and their number depends on the gate voltage. When $V$ is smaller than the interlayer coupling $\gamma_1$, only one gap state connects the valence and conduction band continua, while for $V > \gamma_1$ three gapless states persists. In the latter case, the dispersion $E(k)$ of the middle gap state changes twice its slope. Therefore, the sign of the slope of $E(k)$ of the gap state at the Fermi Energy and $k$ at the cone center changes when the value of $V$ is changed. As a consequence, the sign of the velocity of electrons occupying this state, and thus the direction of the corresponding current, also changes, the effect that potentially could be exploited in nanoelectronic applications based on graphene. 

Additionally, we have demonstrated that the enrgy spectrum of SLG is almost the same as the spectrum of the entire system (except the zigzag-edge states). Moreover, the gap state that at $E_F$ changes its velocity when $V$ changes its value from 0.1 eV to 0.5 eV is localized almost exclusively in this single layer, although it emerges because the adjacent trilayers have different stacking order. Consequently, the one-dimensional current related to this state flows mainly in the single graphene layer. 
We can conclude that the physics of the trilayer system is still present in the  single graphene layer when this system is partially stripped of the outer layers.


\begin{thebibliography}{99}
\bibitem{Ohta_2006} T. Ohta, A. Bostwick, T. Seyller, K. Horn, and E. Rotenberg, Science {\bf 313}, 951 (2006).
\bibitem{Castro_2007} E. V. Castro, K. S. Novoselov, S. V. Morozov, N. M. R. Peres, J. M. B. L. dos Santos, J. Nilsson, F. Guinea, A. K. Geim, and A. H. C. Neto, Phys. Rev. Lett. {\bf 99}, 216802 (2007).
\bibitem{Oostinga_2008} J. B. Oostinga, H. B. Heersche, X. Liu, A. F. Marpurgo, and L. M. K. Vandersypen, Nat. Mater. {\bf 7}, 151 (2008).
\bibitem{Zhang_Nature_2009} Y. Zhang, T.-T. Tang, C. Girit, Z. Hao, M. C. Martin, A. Zettl, M. F. Crommie, Y. R. Shen, and F. Wang, Nature (London) {\bf 459}, 820 (2009).
\bibitem{Szafranek_2010} B. N. Szafranek, D. Schall, M. Otto, D. Neumaier, and H. Kurz, App. Phys. Lett. {\bf 96}, 112103 (2010).
\bibitem{Padilha_2011} J. E. Padilha, M. P. Lima, A. J. R. da Silva, and A. Fazzio, Phys. Rev. B {\bf 84}, 113412 (2011).
\bibitem{Schwierz_2010} F. Schwierz, Nat. Nanotechnol. {\bf 5}, 487 (2010).
\bibitem{Lin_2008} Y.-M. Lin and P. Avouris, Nano Lett. {\bf 8}, 2119 (2008). 
\bibitem{Choi_2010} S.-M. Choi, S.-H.Jhi, and Y.-W. Son, Nano Lett. {\bf 10}, 3486 (2010).
\bibitem{Santos_2012} H. Santos, A. Ayuela, L. Chico, and E. Artacho, Phys. Rev. B {\bf 85}, 245430 (2012).
\bibitem{Zhang_transistor_2018} Q. Zhang, Y. Yaofeng, K. S. Chan, Z. Mu, and J. Li, App. Phys. Express {\bf 1}, 075104 (2018).
\bibitem{Jarillo2018} Y. Cao, V. Fatemi, S. Fang, K. Watanabe, T. Taniguchi, E. Kaxiras, and P. Jarillo-Herrero, Nature {\bf 556}, 43 (2018).
\bibitem{Chen_Nature_2019} E. Chen, A. L. Sharpe, P. Gallagher, I. T. Rosen, E. J. Fox, L. Jiang, B. Lyou, H. Li, K. Watanabe, T. Taniguchi, J. Jung, Z. Shi, D. Goldhaber-Gordon, Y. Zhang, and F. Weng, Nature {\bf 572}, 215 (2019).
\bibitem{Chittari_PRL_2019} B. L. Chittari, G. Chen, Y. Zhang, F. Wang, and J. Jung, Phys. Rev. Lett. {\bf 122}, 016401 (2019).
\bibitem{Yin_PRB_2020} L.-J. Yin, L.-Z. Yang, L. Zhang, Q. Wu, X. Fu, L.-H. Tong, G. Yang, Y. Tian, L. Zhang, and Z. Qin, Phys. Rev. B {\bf 102}, 241403(R) (2020).
\bibitem{Liu_Nature_2020} X. Liu, Z. Hao, E. Khalaf, J. Y. Lee, Y. Ronen, H. Yoo, D. H. Najafabadi, K. Watanabe, T. Taniguchi, A. Vishwanath, and P. Kim, Nature {\bf583}, 221 (2020).
\bibitem{Shen_Nat_Phys_2020} Ch. Shen, Y. Chu, Q. Wu, N. Li, S. Wang, Y. Zhao, J. Tang, J. Liu, J. Tian,K. Watanabe, T. Taniguchi, R. Yang, Z. Y. Meng, D. Shi, O. V. Yazyev, and G. Zhang, Nat. Phys. {\bf 16}, 520 (2020). 
\bibitem{An_SN_2018} M. An,  Q. Deng, Y. Li, H. Song, and M. Su, Superlattices Microstructures {\bf 123}, 172 (2018).
\bibitem{Kishimoto_2016} K. Kishimoto and S. Okada, Jap. J. Appl. Phys., {\bf 55}, 06GF06 (2016).
\bibitem{Menezes_JPB_2015} M.G. Menezes and R. B. Capaz, J. Phys. Condens. Matter {\bf 27}, 335302 (2015).
\bibitem{Jaskolski_2016} W. Jaskolski, M. Pelc, L. Chico, and A. Ayuela, Nanoscale {\bf 8}, 6079 (2016).
\bibitem{Jaskolski_RSC_2019} W. Jaskolski and A. Ayuela, RSC Adv. {\bf 
9}, 42140 (2019).
\bibitem{Yazyev_RPP_2010} O. V. Yazyev, Rep. Prog. Phys. {\bf 73}, 056501 (2010).
\bibitem{Nair_NC_2013} R. R. Nair, I.-L. Tsai, M. Sepioni, J. Keinonen, A. V. Krasheninnikov, A. H. Castro-Neto, M. I. Katsnelson, A. K. Geim, and I. V. Grigorieva, Nat. Commun. {\bf 4}, 1 (2013).
\bibitem{Wu_ACS_Nano_2015} H.-Ch. Wu, A. N. Chaika, T.-W. Huang, A. Syrlybekov, M. Abid, V. Y. Aristov, O. V. Molodsova, S. V. Babenkov, D. Marchenko, J. Sanchez-Barriga, P. S. Mandal, A. Y. Varykhalov, Y. Niu, B. E. Murphy, S. A. Krasnikov, O. Lubben, J. J. Wang, H. Liu, L. Yang, H. Zhang, M. Abid, Y. T. Janabi, S. N. Molotkov, Ch.-R. Chang, and I. Shvets, ACS Nano {\bf 9} 8967 (2015).
\bibitem{Yin_NC_2016} L.-J. Yin, H. Jiang, J.-B. Qiao, and L. He, Nat. Commun. {\bf 7}, 11760 (2016).
\bibitem{Kazemi_APL_2013} A. S. Kazemi, S. Crampin, and A. Ilie, Appl. Phys. Lett. {\bf 102}, 163111 (2013).
\bibitem{Anderson_PRB_2022} P. Anderson, Y.  Huang, Y. Fan, S. Qubbaj, S. Coh, Q. Zhou, and C. Ojeda-Aristizabal, Phys. Rev. B {\bf 105}, L081408 (2022).  
\bibitem{Vaezi_2013} A. Vaezi, Y. Liang, D. H. Ngai, L. Yang, and E.-A. Kim, Phys. Rev. X {\bf 3}, 021018 (2013).
\bibitem{Alden_2013} J. S. Alden, A. W. Tsen, P. Y. Huang, R. Hovden, L. Brown, J. Park, D. A. Muller, and P. L McEuen, Proc. Natl. Acad. Sci. {\bf 110}, 11256 (2013).
\bibitem{San_Jose_2014} P. San-Jose, R. V. Gorbachev, A. K. Geim, K. S. Novoselov, and F. Guinea, Nano Lett. {\bf 14}, 2052 (2014).
\bibitem{Pelc_2015} M. Pelc, W. Jaskolski, A. Ayuela, and L. Chico, Phys. Rev. B {\bf 92}, 085433 (2015).
\bibitem{Jaskolski_2020} W. Jaskolski and G. Sarbicki, Phys. Rev. B {\bf 102}, 035424 (2020).
\bibitem{Jaskolski_2019} W. Jaskolski, Phys. Rev. B {\bf 100}, 035436 (2019).
\bibitem{Jaskolski_2021} W. Jaskolski, Mol. Phys., e2013554, doi.org/10.1080/00268976.2021.2013554.
\bibitem{Jaskolski_2018} W. Jaskolski, M. Pelc, G. W. Bryant, L. Chico, and A. Ayuela, 2D Materials {\bf 5}, 025006 (2018).
\bibitem{Nardelli_1999} M. B. Nardelli, Phys. Reb. B {\bf 60}, 7828 (1999).
\bibitem{Ju_Nature_2015} L. Ju, Z. Shi, N. Nair, Y. Lv, C. Jin, J. Velasco, Jr., C. Ojeda-Aristizabal, H. A. Bechtel, M. C. Martin, A. Zettl, J. Analytis, and F. Wang, Nature (London) {\bf 520}, 650 (2015).
\bibitem{Lin_NL_2013} J. Lin, W. Fang, W. Zhou, A. R. Lupini, J. C. Idrobo, J. Kong, S. J. Pennycook, and S. T. Pantelides, Nano Lett. {\bf 13}, 3262 (2013).
\bibitem{Peeters_PRB_2018} T. L. M. Lane, M. Andelkovic, J. R. Wallbank, L. Covaci, F. M. Peeters, and V. I. Falko, Phys. Rev. B {\bf 97}, 045301 (2018).
\bibitem{Wakabayashi} M. Fujita, K. Wakabayashi, and K, Kusakabe, J. Phys. Soc. Jpn. {\bf 65}, 1920 (1996).
\bibitem{Nakada} K. Nakada, M. Fujita, G. Dresselhaus, and M. S. Dresselhause, Phys. Rev. B {\bf 54}, 17954 (1996).
\bibitem{Jaskolski_2011} W. Jaskolski, A. Ayuela, M. Pelc, H. Santos, and L. Chico, Phys. Rev. B {\bf 83}, 235424 (2011).
\bibitem{nota_zigzag_sublat} Zigzag edge states occur in the case of graphene half-planes (or graphene nanoribbons) with zigzag edges. They have the shape of flat bands beginning at the cone ($k=\frac{2}{3}\pi$) and ending at $k=\pi$. The corresponding wave functions are localized at the sublattice defined by the outermost edge atoms \cite{Wakabayashi,Nakada,Jaskolski_2011}. 
\bibitem{nota_zigzag_degener} Since for each of the outer layers there are two graphene half-planes,  the zigzag edge states are doubly degenerate.
\bibitem{nota_z_05} Here, the zigzag edge states are at $E=\pm 0.5$ eV. They do not couple to the gap states, since they are energetically well separated from the gap. 
\bibitem{Martin_PRL_2008} I. Martin, Y. M. Blanter, and A. F. Morpurgo, Phys. Rev.  Lett. {\bf 100}, 036804 (2008).
\bibitem{Rycerz_2007} A. Rycerz, J. Tworzydlo, and C. W. Beenakker, Nat. Phys. {\bf 3}, 172 (2007).
\bibitem{Kundu_2016} A. Kundu, H. A. Fertig, and B. Seradjeh, Phys. Rev. Lett. {\bf 116}, 016802 (2016).


\end{thebibliography}
\end{document}